\documentclass{article}
\usepackage{spconf,amssymb,amsmath,epsfig}
\usepackage{microtype}
\usepackage{graphicx}
\usepackage{multirow}
\usepackage{booktabs}
\usepackage{amssymb,amsmath,graphicx}
\usepackage{caption}
\usepackage{subcaption}
\usepackage[export]{adjustbox}
\usepackage{moredefs} 
\usepackage{color}

\defcommand{\vec}[1]{\mathbf{#1}} 

\title{An analysis of  incorporating an external language model into a
sequence-to-sequence model}
\name{Anjuli Kannan, Yonghui Wu, Patrick Nguyen, Tara N. Sainath,
\secondlinename{Zhifeng Chen, Rohit Prabhavalkar}
\address{Google, Inc., USA \\
\fontsize{9}{9}\selectfont\ttfamily\upshape
\{anjuli,yonghui,drpng,tsainath,zhifengc,prabhavalkar\}@google.com}}

\begin{document}
\maketitle
\ninept
\begin{abstract}

Attention-based sequence-to-sequence models for automatic speech recognition
jointly train an acoustic model,
language model, and alignment mechanism.  Thus, the language model component
is only trained on transcribed audio-text pairs.  This leads to the use
of {\em shallow fusion} with an external language model at inference time.
Shallow fusion refers to log-linear interpolation with a separately trained
language model at each step of the beam search. In this work,
we investigate the behavior of shallow fusion across a range of conditions:
different types of language models, different decoding units, and
different tasks.
On Google Voice Search, we demonstrate that the use of
shallow fusion with an neural LM with wordpieces yields a 9.1\% relative
word error rate reduction (WERR) over our competitive attention-based
sequence-to-sequence model, obviating the need for second-pass rescoring.
\end{abstract}

\section{Introduction \label{sec:introduction}}

Sequence-to-sequence models have started to gain popularity for automatic
speech recognition (ASR) tasks, particularly for their benefit of folding
various parts of the speech recognition pipeline (i.e., acoustic,
prononcuation and language modeling) into one neural network
\cite{Chan15, Bahdanau16, Graves12, Graves06}.
For example, the Listen, Attend, and Spell (LAS) model jointly learns an encoder,
which serves as an acoustic model, a decoder, which serves as a language
model (LM), and an attention mechanism, which learns alignments.
Recently, a comparison of these different methods showed that
performance still lagged behind a state-of-the-art ASR system with separate
acoustic, pronunciation and language models \cite{RohitSeq17}. The focus of
this paper is to explore a means of making LAS competitive to a conventional
ASR model.

We propose that one reason for the performance degradation could be that the
LAS decoder, which replaces the LM component in a traditional ASR system, is
trained only on transcribed audio-text pairs, which is about 15 million
utterances for the Google Voice Search
task \cite{RohitSeq17}.  In comparison, state-of-the-art LMs are
typically trained on a billion words or more \cite{Rafal16}.  This raises the
question of whether the LAS decoder can learn a strong enough LM from the
training transcripts.  In particular, we posit that in a task like Google
Voice Search, which has a very long tail of queries, the training transcripts
may not sufficiently expose the LAS decoder to rare words and phrases.

However, these words may appear in auxiliary sources of text-only data such as
web documents or news articles, which comprise billions of words.  This work
investigates the impact of training a separate LM on auxiliary text-only data,
and incorporating this model as an additional cost term when decoding a LAS
model.

Several recent works have also investigated the use of LMs with
attention-based models.  \cite{Chan15} demonstrated
significant improvement by rescoring the $n$-best hypotheses produced by LAS
with a $5$-gram LM. \cite{Bahdanau16} extended this idea by
performing log-linear interpolation between LAS and an $n$-gram LM
at each step of the beam search, a method we will henceforth refer to as
\textit{shallow fusion}, following the terminology of \cite{Gulcehre15}.
Shallow fusion was
further studied in \cite{Jan17}, which extended it with use of a coverage
penalty. Both of these works were limited to Wall
Street Journal (WSJ), which, given its scarcity of data, stands to gain
more from an external LM than a large-scale task such as Google Voice Search.
All of these works only investigated $n$-gram LMs, and all focused on
bidirectional models that output graphemes.

The use of an external LM has also been investigated in the context of
training, such that the LAS model could learn when and how to use the LM
\cite{Gulcehre15, Hori17, ColdFusion17}.  These works applied
Recurrent Neural Network (RNN) LMs, but this was largely cited as a means to
make the integration simpler.  None provided a direct comparison of RNN LMs to
$n$-gram LMs.  Further, they were all limited to grapheme systems, with
\cite{Gulcehre15} focused on machine translation.


This work has two goals.  First, we extend the work of \cite{Jan17} by
exploring the behavior of shallow fusion across different
sub-word units and different types of LMs
on a small corpus task. We find that RNN LMs are
more effective at reducing error than $n$-gram LMs, with
the magnitude of this reduction consistent across sub-word units.

The second goal of our work is to explore the behavior of shallow fusion on a
large-scale, large-vocabulary English Voice Search task.  Voice Search has much
more training data than WSJ so it is not clear that the benefits observed on
WSJ should necssarily translate; given sufficient training data,
the LAS decoder may be strong enough to eliminate the effect of any external LM.
Additionally, Voice Search requires a unidirectional model, which has not
previously been studied with shallow fusion.  Ultimately, we find that shallow
fusion with a worpiece-level RNN LM yields a 9.1\% relative WERR on a
competitive unidirectional baseline.

The next two sections will provide more details about the method we use for
integrating the LM and the variants that we compare.  Section~4
describes the setup for our experiments on two different
tasks, and Section~5 provides the results of these experiments.
Finally, in Section~6 we conclude this study.

\section{Shallow fusion with LAS models}

\subsection{Listen, attend, and spell}

\begin{figure} [h!]
\centering
  \includegraphics[scale=0.35]{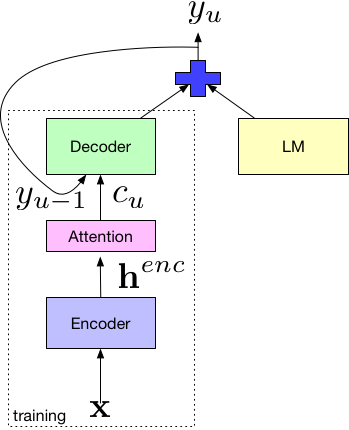}\\
  \caption{The dotted line box shows the basic LAS model, including an encoder,
           attention, and decoder.  In shallow fusion, an external LM is
           incorporated via log-linear interpolation.}
  \label{fig:las_lm}
\end{figure}

As shown inside the dotted line box in Figure~\ref{fig:las_lm},
the LAS model consists of an encoder
(``listen"), an attention mechanism (``attend"), and a decoder (``spell").

The encoder, which is akin to an acoustic model, consists of a stack of
long short-term memory layers (LSTMs)~\cite{HochreiterSchmidhuber97}. These
take as input a sequence of $d$-dimensional feature vectors,
$\mathbf{x} =
(\mathbf{x}_1, \mathbf{x}_2, \cdots, \mathbf{x}_T)$, where $\mathbf{x}_t \in
\mathbb{R}^d$, and produces a higher-order feature representation, denoted
${\mathbf{h}_1^{\text{enc}}}, \cdots, {\mathbf{h}_T^{\text{enc}}}$.

The output of the encoder is passed to an attention mechanism, which
determines which part of the encoder features to attend to in order to predict
each output symbol, effectively performing a dynamic time warping.
The output of the attention mechanism is a single context vector that encodes
this information.

Finally, the decoder is another stack of LSTMs which is conditioned on the
context vector. Given the context vector and the previous prediction $y_{u-1}$
at timestep $u$, the decoder network generates logits
$\mathbf{h}^{\text{dec}_u}$. These are passed through
a softmax to compute a probability distribution
$P(y_u|\mathbf{h}^{\text{dec}_u})$.

The decoder can be thought of as a neural LM conditioned on the acoustic model
output; however, since the LAS model is structured such that the encoder feeds
the decoder, this internal LM can only be trained on audio-text pairs. In the
next section, we will discuss the incorporation of an external LM.

\subsection{Integrating a language model}

Shallow fusion, shown in Figure~\ref{fig:las_lm}, is a method for incorporating
an external LM during inference only.  As the figure shows, only the contents
of the dotted line box are used to train of the LAS model.  At
inference time, however, we perform log-linear interpolation with an LM at each
step of the beam search. In other words, while the objective criterion for
decoding a sequence-to-sequence model typically would be:

\begin{equation}
\mathbf{y^*} = \underset{y}{\arg\max} \log p(y|x)
\end{equation}

\noindent
we instead use the following criterion:

\begin{equation}
\mathbf{y^*} = \underset{y}{\arg\max} \log p(y|x) + \lambda \log p_{LM}(y) + \gamma c(x,y)
\end{equation}

\noindent
where $p_{LM}$ is provided by an LM, and $\lambda$ and $\gamma$ are tuned
on a dev set.  $c(x,y)$ is referred to as a \textit{coverage penalty} and is
designed to penalize incomplete transcripts.  It measures the extent to which
the input frames are ``covered" by the attention weights, computed as:

\begin{equation}
c(x,y) = \sum_{j} \log (\min(\sum_{i} a_{i,j}, 0.5))
\end{equation}

\noindent
where $a_{i,j}$ is attention probability of the $j$th output label $y_j$ on
the $i$th input feature vector $x_i$.  By promoting transcripts which require
attention to more of the audio frames,
the coverage penalty addresses the common sequence-to-sequence failure mode
of assigning high probability to a truncated output sequence \cite{Tu16};
like \cite{Jan17, Yonghui16}, however, we apply this only at decoding time.
The effect of promoting longer transcripts is similar to that of
a length normalization or word insertion reward;  unlike these atlernatives,
however, it is less prone to produce ``babbling", since simply
inserting more tokens while attending to the same frames will not reduce the
coverage penalty.

An alternative method of incorporating an LM  would be to simply rescore the
$n$ best transcripts produced by the beam search, as in~\cite{Chan15}.  Our
initial experiments on the WSJ corpus showed this method provided some
reduction in error,
but not as much as shallow fusion.  This is because the correct prefix may get
pruned by the beam search early on, and not make it into the $n$-best list.

\section{Exploring shallow fusion across tasks, decoding units, and types of language models}

\subsection{Tasks: WSJ vs. Google Voice Search}

This work investigates the impact of shallow fusion on two different tasks.
This is because we hypothesize that there are several task-specific properties
that can affect the relative gain afforded by an external LM:
\begin{itemize}
  \item{\textit{Size of training corpus}, because on a large training corpus
        the LAS decoder will itself be a very strong LM.}
  \item{\textit{Size of vocabulary}, as some of the benefit of an external LM
        may simply be exposure to unseen words and phrases.}
  \item{\textit{Availability of LM training data}, since the LM training data must
        come from the same domain as the task}
\end{itemize}


Our first set of experiments focuses on the WSJ corpus for several reasons.
First, we have a large amount of text-only data also from WSJ, which reduces
the possibility of domain mismatch between the LM and the LAS model.  Second,
given the relatively small size of the WSJ corpus, we see that indeed many
errors in a vanilla LAS model result from a poor LM.  Third, we can use the
standard setup for the training data and vocabulary of the LM, making
comparison to previous works more direct.  Thus WSJ serves as a
useful testbed for measuring the contribution of an external LM.

The small training corpus, however, means that the gains seen on the WSJ task
may not necessarily transfer to a task with a much larger training set.  For
this reason our second set of experiments is done on the Google Voice Search
task.  Two notable properties of Voice Search are that it has a large vocabulary
and it has a very long tail of queries.

\subsection{Decoding Units: Wordpieces vs. Graphemes}

While previous works have only investigated shallow fusion for graphemes, we
extend our study to wordpieces.  Wordpieces \cite{wordpiece_schuster} are
sub-word units that can be as small as a single grapheme
or as large as a complete word.  First, a fixed wordpiece vocabulary is
determined based on frequencies of words in a training corpus.  Once the set of
valid wordpieces is learned, a transcript can be tokenized by choosing the
longest possible component wordpieces in a greedy fashion.

Like graphemes, wordpieces have the advantage that there are
no out-of-vocabulary terms because any word can be decomposed into wordpieces.
(All graphemes are included in the wordpiece vocabulary.)
But wordpieces have the additional benefit that they effectively capture more
context per decoding step than graphemes.  This reduces the length of
dependencies that must be learned by an LM.

For example, the phrase ``the company announced today" consists of 27 graphemes,
which means that a grapheme-level LM (LM-G) would require 27 decoding
steps to output the full phrase; but a wordpiece-level LM (LM-WP) might compose
this phrase as, for example \texttt{the \char`_com pany \char`_announc ed
\char`_today} which would require only 5 steps to output.

As a result, we expect that LM-WP can achieve lower (word-level) perplexity
than LM-G, which could make it more effective in shallow fusion.

\subsection{Language Models: RNNs vs. $n$-gram}

This work further compares shallow fusion across various types of LMs. Previous
works have focused on $n$-gram LMs when applying shallow fusion
\cite{Jan17, Bahdanau16} or RNN LMs for deep or cold fusion
\cite{Gulcehre15, Sennrich15}.  Here we consider both $n$-gram
LMs and RNN LMs \cite{Mikolov2010} for shallow fusion.

There are several reasons that $n$-gram LMs have been preferred in past work.
First, they can incorporate word-level constraints. Since we incorporate the
LM at each step of the beam search, the LM must provide a probability
distribution at the level of the LAS model's decoding unit (either grapheme
or wordpiece). In the case of an RNN LM, this means that we train at the
grapheme or wordpiece level. In the case of an $n$-gram LM, however, there
are two possible setups. The most obvious  is to train the LM at the level of
the decoding unit (grapheme or wordpiece). However, in order to have a strong
grapheme-level LM, it is necessary to train at a very high order, such as
20-gram, to capture at least a few words worth of context.  Following
\cite{Bahdanau16}, an alternative is to train the LM at the word level, and
then, using the Weighted Finite State Transducer framework \cite{Mohri2002, Allauzen2007},
compose it with a ``speller'' which breaks each word into its component units
(graphemes or wordpieces).  In this way, we can still get a probability
distribution at the unit level, while incorporating the knowledge of a
word-level LM.

Furthermore, this latter setup implicitly introduces a dictionary.  In a
task like WSJ, the baseline model has a relatively weak decoder, so it will
frequently output sequences of graphemes which do not comprise English words.
The dictionary constraints imposed by the $n$-gram LM can be helpful to prune
these out.

Finally, in a task like Google Voice Search, there are many sources of
data that can potentially be useful in an external LM.  We can use
Bayesian interpolation to combine $n$-gram LMs trained individually on each of
these domains, optimizing the interpolation weights against WER on a dev set
\cite{Allauzen2011}.  Currently this sort of technique only exists for
$n$-gram LMs.

Despite all these advantages of $n$-gram LMs, recent literature has shown that
state-of-the-art RNN LMs have a significantly lower perplexity than $n$-gram LMs
on the 1 billion word benchmark, particularly on rare words \cite{Rafal16}.
Thus we hypothesize that they should also provide a greater reduction in
error when used in shallow fusion.  Furthermore,
given enough training data, as we have in the Google Voice Search task, we
suggest that the introduction of the dictionary may not be necessary; in fact,
it may be limiting to the model since the LAS model can actually ``sound out"
words that it has never seen before but which are spelled phonetically. Though
the techniques of Bayesian interpolation and incorporating dictionary
constraints currently apply only the $n$-gram models, we posit
that analogous methods should be possible for RNN LMs, and identify these as
areas for future work.

\section{Experimental details \label{sec:experiments}}

\subsection{Wall Street Journal}

Our experiments are conducted on two tasks.  The first is the WSJ dataset.
Following the setup in \cite{Jan17}, we train on \textit{si284}, validate on
\textit{dev93} and evaluate on \textit{eval92}.

For grapheme experiments, our baseline model is a LAS model with 3
convolutional layers and a convolutional LSTM layer, followed by 3
bidirectional~\cite{Schuster97} LSTM layers.
The output vocabulary is 72 graphemes.
Temporal label smoothing is applied as described in \cite{Jan17}.
For wordpiece experiments, our baseline model has the same
architecture as the grapheme model, except that the output vocabulary has 1,024
wordpieces and no label smoothing is applied because label smoothing resulted in
a weaker model.  Instead, L2 regularization is used.

The external LMs are trained using the WSJ text corpus  and extended vocabulary
provided in the Kaldi WSJ s5 recipe \cite{Povey2011}.  The RNN LMs consist
of two LSTM layers of 512 hidden units.  The word-~, grapheme-~, and
wordpiece-level $n$-gram LMs are all trained with Katz smoothing and pruned to
between 15M and 20M $n$-grams.  The word-level LM is composed with a speller
to decode at the grapheme or wordpiece level.

\subsection{Google Voice Search}

The second task is a $\sim$12,500 hour training set consisting of
15M English utterances.
The training utterances are anonymized and hand-transcribed, and are
representative of Google's Voice Search traffic.
This data set is created by artificially corrupting clean utterances using a
room simulator, adding varying degrees of noise and reverberation such that the
overall SNR is between 0dB and 30dB, with an average SNR of 12dB.
The noise sources are from YouTube and daily life noisy environmental
recordings. We report results on two sets of $\sim$14,800 anonymized,
hand-transcribed Voice Search utterances each, extracted from Google traffic.

The baseline model for Voice Search experiments has an encoder consisting of
5 unidirectional LSTM layers of 1,400 units each, a decoder consisting of
2 LSTM layers with 1,024 hidden units each, and a multi-headed attention
mechanism \cite{Vaswani17}.  We use a unidirectional encoder because the Voice
Search task requires a streaming model.

All experiments use 80-dimensional log-mel features, computed with a 25-ms
window and shifted every 10ms. Similar to~\cite{Hasim15, Golan16}, at the
current frame, $t$, these features are stacked with 3 frames to the left and
downsampled to a 30ms frame rate.  The models are trained with the
cross-entropy criterion, using asynchronous stochastic gradient descent
optimization in TensorFlow~\cite{AbadiAgarwalBarhamEtAl15}.

Our text dataset consists of billions of sentences from
several sources: untranscribed anonymized Voice Search queries, untranscribed
anonymized voice dictation queries, anonymized typed queries from Google Search,
as well as the transcribed training utterances mentioned above.
The production LMs denoted as \textsc{prodlm1} and \textsc{prodlm1} are
both 5-gram LMs with a vocabulary of 4M.  \textsc{prodlm1} is
constructed as a Bayesian-interpolated mixture of LMs trained on the individual
data sources \cite{Allauzen2011}, while \textsc{prodlm2} is trained on all
data.  Following \cite{Rao17}, the RNN LM is trained
on about half a billion sentences sampled from the full pool data. It
consists of two LSTM layers of 2,048 units each.

\section{Results \label{sec:results}}

\subsection{Comparing LMs for shallow fusion}

We begin by comparing three types of LMs in the context of shallow fusion with
the LAS grapheme model \textsc{las-g}  on the WSJ task: (1) an RNN LM trained on graphemes
(\textsc{rnn-g}), (2) a 20-gram LM trained on graphemes (\textsc{20-gram-g}),
and (3) a 3-gram LM trained on words and composed with a speller
(\textsc{3-gram-w}).

Comparing these, we see that \textsc{3-gram-w} barely outperforms
\textsc{20-gram-g}.
This shows that, given the same amount of context, having word constraints and
an implicit dictionary has only a slight benefit.
\textsc{rnn-g}, however, outperforms both of
the $n$-gram LMs, suggesting while the word constraints may help,
they are insufficient to make
up the gap between RNN LMs and $n$-gram LMs.  One opportunity for future
work would be incorporating word constraints into \textsc{rnn-g}.

\begin{table} [h!]
\centering
\begin{tabular}{|c|c|c|c|} \hline
System & Dev & Test \\ \hline
\textsc{las-g}  &  13.0 &  10.3  \\
\textsc{las-g}  + \textsc{20-gram-g}   & 10.3 &  7.7  \\
\textsc{las-g}  + \textsc{3-gram-w}   & 10.0 &  7.6  \\
\textsc{las-g}  + \textsc{rnn-g} & 9.3 & 6.9 \\ \hline

\end{tabular}
\vspace{-0.1 in}
\caption{WER of \textsc{las-g} fused with various LMs.  While word
constraints do help the $n$-gram LM, \textsc{rnn-g} performs even better.}
\vspace{-0.1 in}
\label{table:wsj_char}
\end{table}

\subsection{Extending shallow fusion to wordpiece models}

Next, we perform a comparison for \textsc{las-wp}. Since we have shown that
word constraints are helpful for sub-word-level $n$-gram LMs, we limit our
comparison to just two LMs: (1) an RNN LM trained on wordpieces
(\textsc{rnn-wp}), and (2) a 3-gram LM trained on words and composed with a speller
(\textsc{3-gram-w}).

As Table \ref{table:wsj_wpm} shows, we see the same trend on \textsc{las-wp},
with \textsc{rnn-wp} significantly better than \textsc{3-gram-w}.  However, it
should be noted that the baseline \textsc{las-wp} is worse than \textsc{las-g}.
This is likely due to the small amount of data being insufficient to train
the large number of additional parameters: we found that the larger we made the
wordpiece vocabulary, the worse the model became.  As a result of this difference, the
LM results for \textsc{las-wp}  are not directly comparable to the LM results
for \textsc{las-g}.  The main observation we make is that the RNN performs best
in both cases, with the relative improvement being roughly consistent for both
graphemes and wordpieces.

\begin{table} [h!]
\centering
\begin{tabular}{|c|c|c|c|} \hline
System & Dev & Test \\ \hline
\textsc{las-wp}  &  15.7 &  12.3  \\
\textsc{las-wp}  + \textsc{3-gram-w}   & 12.9 &  9.3  \\
\textsc{las-wp}  + \textsc{rnn-wp} & 11.5 & 8.2 \\ \hline
\end{tabular}
\vspace{-0.1 in}
\caption{WER of \textsc{las-wp}  combined with various LMs on WSJ.
\textsc{rnn-wp} again performs best.}
\vspace{-0.1 in}
\label{table:wsj_wpm}
\end{table}

\subsection{Scaling up to Voice Search}

We now turn to the Voice Search task. First, since we have an abundance of
training data, we see in the first two lines of
Table~\ref{table:vs_lm} that the wordpiece model
(\textsc{las-wp}) is now comparable with the grapheme model (\textsc{las-g}).
Thus our analysis here is limited to \textsc{las-wp}.

In the traditional HMM/CTC-based system, the decoding proceeds in two passes:
the first pass uses a small $n$-gram LM (\textsc{prodlm1}), which fits in memory
and minimizes the search space to meet real-time requirements. The first pass
generates an N-best list which we rescore with a much larger $n$-gram LM
(\textsc{prodlm2}) \cite{Allauzen2011}.
In the third and fourth lines of Table~\ref{table:vs_lm} we see the results of
applying the production LMs to the LAS model with shallow fusion: the LM
inherent in LAS is quite competitive, but there is a small gain from
the highly-pruned \textsc{prodlm1}. The much larger
\textsc{prodlm2}, despite being 40x larger, provides only slightly more
improvement.  In addition, \textsc{prodlm2} is 80GB and must be run on
multiple servers. This is operationally unwieldy and cannot be efficiently
integrated with low latency during the first pass.

On the other hand, while computationally expensive, RNN LMs are known to be
more compact than their $n$-gram counterparts.
In line 5 of Table~\ref{table:vs_lm}, \textsc{las-wp}  + \textsc{rnn-wp},
we show that the shallow fusion of LAS with  \textsc{rnn-wp} provides
an even greater benefit than \textsc{prodlm2}.  Its much lower memory footprint
(1.1 GB) allows it to fit in the first pass. We then rescore the system with
\textsc{prodlm2} (as \textsc{las-wp}  + \textsc{rnn-wp} + \textsc{prodlm2}).
This yields no further gain, showing that we have obviated the need for a
second-pass rescoring at all.

Thus, as with WSJ, we see that \textsc{rnn-wp}
more effectively encodes the LM information compared to the $n$-gram model.
In addition, \textsc{rnn-wp} is 1.5\% the size of \textsc{prodlm2}, and also
enjoys the additional benefit of not having out-of-vocabulary words since it
is trained on wordpieces.  Note that both \textsc{prodlm1} and
\textsc{prodlm2} are interpolated across several data-source-specific LMs,
while \textsc{rnn-wp} uses ad hoc mixing
weights for the various data sources.  Investigating a more principled method
of mixing the data sources for \textsc{rnn-wp} is an opportunity for future
work.

\begin{table} [h!]
\centering
\begin{tabular}{|c|c|c|c|c|} \hline
System & Dev & Test & LM size  \\ \hline
\textsc{las-g}  & 9.5  & 7.7  & 0GB \\
\textsc{las-wp}  & 9.2  & 7.7  & 0GB \\ \hline
\textsc{las-wp}  + \textsc{prodlm1} & 8.8 & 7.4 & 2GB \\
\textsc{las-wp}  + \textsc{prodlm2} & 8.7 & 7.2 & 80GB\\\hline
\textsc{las-wp}  + \textsc{rnn-wp} & 8.4 & 7.0 & 1.1GB\\
\textsc{las-wp}  + \textsc{rnn-wp} + \textsc{prodlm2} & 8.4 & 7.0 & 81.1GB\\ \hline
\end{tabular}
\vspace{-0.1 in}
\caption{WER of shallow fusion of LAS with production $n$-gram LMs and an RNN LM. The RNN LM captures all the benefits of \textsc{prodlm2}
  in a compact form.
}
\vspace{-0.1 in}
\label{table:vs_lm}
\end{table}

\section{Conclusions \label{sec:conclusions}}

In this work we investigated the technique of shallow fusion, in which an
external LM is used to augment a LAS model at inference time.
We demonstrated that on the small WSJ task, an RNN LM yielded
greater improvement than an $n$-gram LM, and the gains were consistent across
graphemes and wordpieces.
On the much larger Voice Search task, we showed that the  decoder LM inherent in
LAS is already very competitive, yielding little benefit from shallow fusion with
the first-pass production LM.
However, we found that shallow fusion with an RNN LM provided greater benefit.
In fact, with 9.1\% relative WERR on a competitive
unidirectional system, it eliminated the need for a second pass rescoring,
despite being 70 times smaller than the second pass LM.

\section{Acknowledgements}
The authors would like to thank Jan Chorowski, Navdeep Jaitly, Shankar Kumar,
Kanishka Rao, Brian Roark, and David Rybach, for helpful discussions.
\bibliographystyle{IEEEbib}
\bibliography{main}
\end{document}